# Analysis of Point-contact Andreev Reflection Spectra in Spin Polarization Measurements

G.T. Woods[1,4], R, J. Soulen Jr.[1], I. I. Mazin.[1], B. Nadgorny[2], M. S. Osofsky[1], J. Sanders[3], H. Srikanth[3], W. F. Egelhoff and R. Datla[4]

[1]*Naval Research Laboratory, Washington, DC 20375*

[2]*Dept. of Physics and Astronomy, Wayne State University, Detroit, MI 48201*

[3]*Dept. of Physics, University of South Florida, Tampa, FL.33620*

[4]*National Institute of Standards and Technology, Gaithersburg, Md.*

## Abstract

We present a systematic analysis of point-contact Andreev reflection (PCAR) spectra for ferromagnetic materials, using both modeling and experimental data. We emphasize the importance of consistent data analysis to avoid possible misinterpretation of the data. We consider the relationship between ballistic and diffusive transport, the effect of different transport regimes on spin polarization measurements, and the importance of unambiguous identification of the type of transport regime. We find that in a realistic parameter range, the analysis of PCAR spectra of purely diffusive character by a ballistic model yield approximately the same (within ~3%) values of the spin polarization and the barrier strength $Z$ larger by ~ 0.5-0.6. We also consider the dependence of polarization values on $Z$, and have shown by simple modeling that letting the superconducting gap vary as an adjustable parameter can result in a spurious dependence of the spin-polarization $P_c$ on $Z$. At the same time we analyzed the effects of finite $Z$ on the apparent value of $P_c$ measured by the PCAR technique, using a large number of examples from both our own measurements and from the literature. We conclude that there is a system-dependent variation in $P_c$ ($Z$), presumably due to spin-flip scattering at the interface. However, the exact type of this dependence is hard to determine with any statistical certainty.

Spin-polarized ferromagnetic materials are set to play a key role in the next-generation electronic devices, based on the electron's spin rather than charge.[1] The performance of many of these *spintronics* devices improves dramatically as the spin polarization, *P*, of the ferromagnetic material increases[1]. Particular attention has focused on the so-called 'half-metals', in which the electrons responsible for the metallic transport all have the same spin (either spin up or spin down), while the electrons with the opposite spin are insulating. Half-metals have the maximum attainable value of spin polarization (100%). Most of the experimental studies to determine *P* have been carried out by the spin-dependent tunnelling technique, pioneered by Tedrow and Meservey.[2] This method requires the material of interest to be fabricated as part of a ferromagnet/superconductor tunnel junction, in which the superconducting density of states is then Zeeman-split by the application of a magnetic field of several Tesla. The other conventional technique is spin-resolved photoemission, which measures the spin of the electrons emitted from a region close to the surface of the ferromagnet of the order of 5–20 A, and thus is quite surface-sensitive.[3]

The point contact Andreev reflection (PCAR)[4,5] technique, which is considerably easier to put into practice, serves to expedite and widen the searches for many new materials that are too difficult to incorporate into tunnel junctions. PCAR is a technique in which the conductance ($G \equiv dI/dV$) is measured for an electrical point contact with little or no tunneling barrier established between a superconducting tip and a ferromagnetic counter-electrode (or *vice versa*). The presence of spin-polarized current in the ferromagnet alters the conductance of the contact in a known way, giving rise to a new technique to determine $P_c$ (the spin polarization measured by PCAR). This method offers several advantages. With no restrictions on the sample geometry, one can avoid complex fabrication steps. In addition, the PCAR method has excellent energy resolution (~0.1 meV), and does not require an applied magnetic field. Perhaps the only significant disadvantage of PCAR, compared to other techniques (unless it is done *in situ*), comes from a possible surface modification, due to uncontrolled surface oxides, or other chemical reactions on the surface of both the ferromagnet and the superconductor. The effect of this surface modification on $P_c$ is impossible to quantify, but, this drawback notwithstanding, the results obtained by the PCAR technique are usually in good agreement with values obtained by other methods where the surface oxidation is better controlled (Ref. 2). A plausible explanation for the success of the PCAR method is that the fragile surface oxide layer is usually penetrated as the point contact is established.

PCAR spectra to date have been typically analyzed using a modification of the Blonder-Tinkham-Klapwijk (BTK) model[6], to include a spin polarization of the metal.[4,7] It is a weak coupling theory that combines all interface effects into a single dimensionless parameter, $Z$, which does not necessarily correspond to any physical parameter characterizing the interface barrier. Recently, the applicability of the BTK and the modified BTK formulas to the spin polarization measurements has been questioned.[8] Undoubtedly, the BTK theory neglects some of the delicate surface phenomena, such as the presence of surface states and the effects of lattice relaxation. The theory also makes assumptions the validity of which are difficult to evaluate, such as the $\delta$–functional form for the barrier, the step-function shape of the voltage drop across the barrier, and lateral momentum conservation. In addition, the modified BTK model also assumes spin-independent barriers.[4,7] Xia et al (Ref. 8) performed advanced LDA calculations for Andreev transport across realistic interfaces. These calculations took care of some, but not of all the issues listed above. However, they were not able to satisfactorily describe experimental curves[5], including the contacts with non-magnetic metals, while the modified BTK formulas, treating $Z$ as an adjustable parameter, provide an excellent description of the same curves. Although the formalism of Ref. 6 is based on a derivation where a $\delta$-shaped barrier is assumed, $Z$ actually incorporates more physics than just the strength of the $\delta$-function, and, therefore, the formalism works much better than could have been expected.[9] That $Z$ is not the real barrier strength in actual measurements is emphasized by the fact that sometimes the BTK model fits experimental curves surprisingly well with $Z = 0$, although formally, due to the Fermi velocity mismatch between the metal and the superconductor, there always exists some minimal non-zero $Z$. Probably the most illustrative case is that of the colossal magnetoresistance material $La_{1-x}Sr_xMnO_3$ ($LSMO$) [10], where because of the large disparity between the Fermi velocities of the majority and minority spin electrons, at least in one spin channel the Fermi velocity mismatch should be very large. [11]

Another important question which was raised in connection with the BTK formalism is the difference between the ballistic and diffusive transport, defined by the ratio of the mean free path *l* of the electrons and the contact diameter, *d*. In general, there are *three* possible types of transport in a PCAR experiment: ballistic ($l>>d$), diffusive ($l<<d$) and intermediate ($l \sim d$). One way of estimating the mean-free path is from the Drude formula, using a measured value for $\sigma$ ($\sigma = ne^2l/mv_F$), where ($n/m$) and $v_F$ can be calculated from the band structure. The diameter of the contact *d* can then be calculated employing the equation for the junction (contact) resistance [12]

$$R_N = R_0(1+Z^2) \approx (4\rho l/3\pi d^2 + \rho/2d)(1+Z^2), \tag{1}$$

where the first term in the expression for $R_o$ is the so-called Sharvin resistance[13] for ballistic contacts, while the second is the Maxwell resistance [14] for diffusive transport. To determine *d*, $R_N$ must be measured and *Z* obtained by analyzing the conductance curves. Alternatively, one can estimate the size of the contact *d* and the mean free path independently (and potentially more accurately) by experimentally measuring the contact resistance in a broad temperature range, which, however, requires high thermal stability of the contacts.[15] In many cases, it is hard to avoid fairly large uncertainties in making such estimates and thus it is often difficult to establish the exact transport regime for the junction conduction. Additionally, the ratio of *l/d* can be often close to one, thus indicating the transport regime in the intermediate region. The applicability of the ballistic theory for the data obtained in this regime, especially given the uncertainty in estimating *l* and *d*, may seem problematic.

According to Ref. 7, it is possible to extend the BTK theory onto the diffuse limit. There is an issue as to which model should be applied to a given set of data. Moreover, no theory has yet been developed for the intermediate case. It is believed that both the ballistic and diffusive formalisms will yield approximately the same value for $P_c$, albeit with different *Z* values, when analyzing the same spectra. If so, and assuming that the behavior for the intermediate case is bracketed by the ballistic and diffusive limits, then it does not really matter which regime applies to a particular junction in a particular experiment, as long as the value of *Z* is not of interest (one should, however, be aware that the actual spin polarization of a material may be different for the ballistic and diffusive transports[19]). This fact may explain why several different groups, exercising no particular control over the transport regime for their point contacts, and using only the ballistic model for the analysis (i.e., implicitly assuming the ballistic regime), still obtained comparable results for $P_c$ for the same materials. Earlier attempts to analyze the same experimental spectra with both diffusive and ballistic formulas seemed to indicate that the resulting polarizations are very close[11]; however, no systematic tests of this assumption have been performed.

In this article, we will present an analysis of PCAR spectra in both the ballistic and diffusive limits for several ferromagnetic oxides, mainly *CrO₂*, *SrRuO₃* (SRO) and *LSMO*, in order to illustrate some possible caveats in applying the modified BTK formulas to real materials. Firstly, we will discuss the differences between the diffusive and the ballistic models. Secondly, we will consider the sources of possible systematic errors, when analyzing $G(V)$ curves. Specifically, we will discuss the effect of using the superconducting gap, $\Delta$, as a variable parameter on the extracted value of $P_c$ and show how, with the inclusion of the additional spreading resistance $R_s$ of the sample at a given experimental temperature, the effects of $\Delta$ on $P_c$ can be eliminated. Finally, we will consider the possible $P_c$ vs. $Z$ dependence seen frequently in PCAR spectra by performing a systematic analysis of a large number of different experiments. It has been argued that the functional dependence of $P_c(Z)$ is quadratic[10,16,17], or exponential.[18] Using statistical analysis, we will show that either exponential, quadratic, or linear dependence has no apparent advantages over the others.

As we have mentioned above, two different models, ballistic and diffusive, may be used to extract values of $P_c$ from the data for $G(V)$. Both models separate the current at the N/S interface into spin-polarized and non-polarized contributions, and give the expressions for $G(V)$ for the two transport regimes in terms of the superconducting order parameter $\Delta$, the bias voltage, and the interfacial barrier strength $Z$. Table 1[7] shows the equations for the total current at the interface. In addition, the equations that describe the conductance also contain pre-factors in terms of the density of states $N$ at the Fermi level and the Fermi velocity $v$ of both majority and minority spins. For the ballistic case, the pre-factors are $<Nv_{\uparrow,\downarrow}>$, while those for the diffusive are $<Nv^2_{\uparrow,\downarrow}>$. In practice, for both models one also needs to include corrections for the spreading resistance of the sample $R_s$, the additional resistance of the sample between the junction and one of the electrical contacts in a typical four-probe measurement scheme. The presence of $R_s$ results in the shift of the apparent position of the coherence peak $G(V)$ from $V \approx \Delta$ to larger voltages and in the change of the observed zero bias conductance value. These effects have to be always taken into account, unless $R_s$ is much lower than the junction resistance, which is usually the case only for bulk samples or highly conductive films. $\Delta$ and $R_s$ can be used as fitting parameters or can be determined experimentally, as was done in this paper.

First we pose the following question: *If an experimental PCAR spectrum with zero or finite Z is obtained in the diffusive regime, can one apply the ballistic, rather than diffusive model to analyze it?* Furthermore, if this is possible, how will the values of the parameters ($P_c$ and $Z$) compare? To answer this question, we first generated a large number of $G(V)$ curves for a hypothetical superconductor using the diffusive model with given values of $P_d$ and $Z_d$ (diffusive spin-polarization and barrier strength respectively). We then analyzed these curves using the ballistic model to obtain "ballistic" values of the spin-polarization $P_b$ and the barrier strength $Z_b$. The values of $T$, $\Delta$ and $R_s$ were 0.1 or 1.5 K, 1 meV and 1 $\Omega$ respectively and these values were kept throughout. The results for this procedure are shown in Fig. 1a,b where we plot $P_b - P_d$ and $Z_b - Z_d$ vs. $P_d$ for values of $Z_d$ = 0.0 and 0.75. The two main points illustrated by Fig. 1 are that: (1) Although this procedure tends to overestimate $P_c$ for small polarizations, and slightly underestimate it for $P_c$~60-80%, potential error introduced by applying the ballistic formulas to the diffusive contacts is negligible, less than ±3% in absolute value, for most of the spin polarization range; (2) Whereas the obtained values of $Z$ differ significantly for the two models, a comparison between the values of $Z_b$ and $Z_d$ for all fits, showed that $Z_b$ is always greater; for small $Z_d$ and $P_d$ the difference is 0.5-0.6. This is one of the illustrations of the hidden power of the BTK model: in our diffusive case, where we do have a δ-functional barrier with a *known* strength, *plus* another physical effect, not accounted for in the original BTK formalism, we see that the single parameter $Z$ absorbs all this additional physics, producing practically the same values of the spin polarization.

The second question we pose is: *What is the effect of changing the value of the superconducting gap on the extracted values of the spin polarization?* As we have mentioned above, it has become a rather common practice in the PCAR studies to take a succession of $G(V)$ curves for different point contacts and to analyze each one of them to obtain the values of $P_c$ and Z. Thereafter one plots $P_c$ vs. Z, which is then extrapolated to $Z = 0$ to obtain an "intrinsic" value of $P_c$ for the system.[10,16-18] . However, quite often the coherence peak is displaced from its theoretical position near the bulk superconducting gap. This effect can have two different causes: variation of the superconducting gap near the interface or the presence of the spreading resistance, $R_s$. In fact, both $\Delta$ and $R_s$ in every experiment should be uniquely determined. $R_s$ can be measured independently, whereas $\Delta$ can be inferred from $T_c$ using the BCS model. However, in many cases the analysis is done using $\Delta$ as an *adjustable parameter*, which, as we will show below, can strongly affect the values of the spin-polarization. Varying $\Delta$ is related to varying $R_s$, in the sense that both shift the apparent coherence peak from its BCS value, albeit in the opposite directions.

To illustrate the relationship between $\Delta$ and $R_s$, we plot several theoretical curves for the same $\Delta$, spin-polarization and Z, but different $R_s$ in Fig. 2 (inset). This imitates an experimental situation when several contacts with different $R_s$ are measured. It then appears that we can describe the same set of curves with the *same $R_s$* if we allow the gap to vary from curve to curve. Fig. 2 shows the resulting dependence of $\Delta$ on $R_s$. Importantly, now the two other parameters, P and Z are also different for different curves. In other words, by analyzing experimental data collected with different $R_s$ as if they all had the same $R_s$ (or no spreading resistance at all), the wrong $\Delta$ is obtained and this error propagates into the value of the spin-polarization (see Fig. 3).

Obviously, spurious dependencies appear in such a case for all three quantities: $\Delta$, Z and P. This can be easily mistaken as a dependence of P on Z, as we show in Fig. 3. There we used a *single* point contact spectrum of *Sn/LSMO* (contact #10) for a temperature of approximately $T = 1.75$ K. If we assume different values for $R_s$, the conductance curves $G(V)$ as a function of the voltage at *the point contact* will be different (see the inset in Fig.3). We then analyzed the resulting curves using the standard BTK formulas and find a *different* value for $P_c$ for each curve. As we can see from the plot, ~ 1% error in $\Delta$ corresponds to ~ 1% error in $P_c$. Therefore, it is always desirable to evaluate the gap and the spreading resistance separately. If this is not possible, it may be prudent to fix the value of $\Delta$, rather then let it vary as an additional parameter. However, if the apparent position of the coherence peak in the raw spectra is shifted to significantly smaller values than the bulk gap, it may be an indication of a surface suppression of the order parameter, in which case more elaborated models are needed.

That brings us to another important point, namely, whether or not the dependence of $P_c$ on the value of $Z$, often reported in the literature, is real. As one can see from Fig.3, in this case there is a clear correlation not only between $\Delta$ and $P_c$, but also between the value of the Z-parameter and $P_c$, as $Z$, in turn monotonically changes with the gap. On the other hand, we *know* from the onset, that the actual data in Fig.3 corresponds to just one $Z$ (the same way it corresponds to a single value of $R_s$), so the "observed" Z-dependence is utterly spurious. Note that the limiting value of $P_c$ at $Z =0$ in this case is not necessarily the "intrinsic" spin polarization, as both the gap value and $R_s$ corresponding to this Z may be incorrect. However, we don't want to leave the reader with a conclusion that all of the observed $P_c$ (Z) dependencies are artifacts, and, as we will show below, in a number of cases we did observe this dependence, in spite of all possible precautions in analyzing the data.

To further discuss the two models, we present $Pb/CrO_2$ data, which is analyzed in both the ballistic and diffusive limits. The (100) $CrO_2$ films used in this study were made by the Chemical Vapor Deposition (CVD) method described extensively elsewhere.[20] The measurements of these surface-sensitive samples were done immediately after the film deposition in order to avoid any film degradation. The measurements with $Pb$ and $Sn$ contacts were performed in a liquid He bath at temperatures between 4.2 K and 1.5 K using the technique described elsewhere.[4,11]

Plotted in Fig. 4 are the experimental $G$ (V)/ $G_N$ spectra of $CrO_2$, which were fitted using the (a) ballistic model and (b) the diffusive model for a high-Z contact (#9) and for a low-Z contact (#4). Displayed along with the spectra are the fitted values of $P_c$ and $Z$. Each spectrum was obtained at a temperature of approximately 1.75 K. We calculated values of $\Delta$ for this temperature using the BCS approximation ($\Delta$ = 1.2 meV and 0.55 meV for $Pb$ and $Sn$ respectively) and kept them constant throughout the analysis. We have also used the values of the experimentally determined $R_s$, and made sure that it is the same for all the contacts measured in the same geometry. Using this procedure, both models gave nearly the same value of $P_c$ for contact #4 ($P_b = 0.81 \pm 0.03$, $P_d = 0.85 \pm 0.03$), and for contact #9 ($P_b = 0.44 \pm 0.03$, $P_d = 0.47 \pm 0.03$) as well as all other analyzed PCAR spectra in this experiment (see Fig. 5).

As mentioned previously, there should be no correlation between $P_c$ and $Z$ in the BTK formalism. We have also shown above that some of the $P_c(Z)$ dependencies may be caused by systematic errors due to inconsistent analysis of the data. Nevertheless, we have observed such a correlation in at least some of the material systems, in which this correlation had been previously reported, most notoriously in $CrO_2$. To illustrate how $Z$ affects $P_c$ in our spectra, we plot $P_c$ vs. $Z$ for our $Pb/CrO_2$ data in Fig. 5. We first use the ballistic formula, and obtain polarizations between 20 and 85%, with $Z$ between 0.55 and 1.45. $P_c$ indeed decreases with increasing $Z$ in agreement with other studies of this material.[16] Despite the fact that there are no theoretical arguments for a linear relationship between $P_c$ and $Z$, the fitted values in Fig. 5 show a fairly good linear dependence. However, if we extrapolate to $Z = 0$ linearly, we obtain $1.13 \pm 0.06$, which is unphysical. In Ref. 16 a quadratic dependence of $P_c(Z)$ for $CrO_2$ was proposed. While this is also hard to justify theoretically, a quadratic extrapolation gives $P_c(Z = 0) = 1.05 \pm 0.29$. This result gives a more realistic number for $P_c(Z = 0)$, closer to the theoretical value for this system[21], but with a larger degree of uncertainty, which indicates that there are no statistical arguments for using a quadratic dependency for this set of data. This is, of course, related to the fact that we were not able to collect any data for this sample that could be described by the ballistic model with $Z<0.5$. On the other hand, the same spectra can be fitted by the diffusive model with practically the same polarization values, but with $Z$ varying from 0 to 1.1. So, the diffusive model for ($Z \cong 0$) yields $P_c = 0.85 \pm 0.03$ without any extrapolation. Thus, if we were dealing with an unknown material we would have a dilemma: to either use the ballistic model and quadratic extrapolation to $Z=0$, but with a large uncertainty, or the diffusive model without extrapolation and thus with a smaller value of $P_c$ (and, if the linear extrapolation would not yield $P_c > 100\%$, we would have to think about this alternative as well). In this specific case, as the film was of relatively low quality and with high residual resistivity, it is likely that our sample does, indeed, have $P_c<1$. In other words, the correct value of $P_c$ in this case is probably the one given by the diffusive model. The fact that we were not able to obtain any spectra with $Z_b$<0.5, which is the minimal $Z_b$ that can be obtained in the diffusive regime (Fig. 1), can serve as a red flag suggesting that we are, indeed, in the diffusive regime. On the other hand, if in an experiment $Z_b$<0.5 is observed, this is a good indication that ballistic formulas should be applied, with a subsequent extrapolation.

As stated above, there is no theory that suggests $P_c$ $(Z)$ should be either linear or quadratic. However, Kant et al. (Ref. 18) proposed that $P_c$ could be written as

$$P \approx P_0 \exp(-2\alpha\psi Z^2),$$

where $P_0$ is the intrinsic value of the spin-polarization, $\alpha$ is defined as the spin-flip scattering probability and $\psi$ is the ratio of the forward and backward scattering probabilities. The above equation from Ref. 18 was derived by a relatively elaborated procedure. However, its physical meaning is very simple: in their model, $Z^2$ is derived from multiple scattering within the interface region (it is noteworthy that this assumption is applicable only for diffusive contacts though the authors apply it in the ballistic case). Obviously, $Z^2$ is proportional to the number of collisions and therefore to the ratio $d/l$. On the other hand, a natural (but not always correct) interpretation of the polarization suppression with $Z$ is spin-flip scattering by impurities in the interface. This is also proportional to the number of scattering events, albeit that only a small fraction of scattering results in a spin flip. This immediately leads to Eq. 2, where $\alpha \ll 1$ is of the order of $l/l_{sd}$, where $l_{sd}$ is the spin diffusion length. Interestingly, even when actual data can be fitted by an exponential formula, the product $\alpha\psi$ both in Ref. 18 and our own similar calculations (See Table 2) is of the order of, and not much smaller than one, which simply reflects the fact that the assumption of a diffusive regime, implicitly used in the derivation, does not hold. On the other hand, it is obvious that for poor contacts with large $Z$ and strong spin-flip scattering the apparent value for $P_c$ should tend to zero. Furthermore, since the total contact resistance $R_N$ in the BTK model is proportional to (1 + $Z^2$), it is natural to assume that in many cases the spin-flip scattering, whether from impurities or not, depend on $Z^2$, and not on $Z$. Therefore, the exponential function, which smoothly interpolates between the two limits, may have some validity. Nonetheless, there is no significant improvement in using Eq. 2 over a quadratic or even a linear dependence (cf. the values of the $\chi^2$ criterion for the three fits as shown in Table 2 and plotted in Fig. 6. For all materials the three $\chi^2$ values are very close, which indicates that all three extrapolations are of comparable statistical quality).

In summary, we have discussed an analysis of PCAR spectra using the ballistic and diffusive models. By careful analysis of the PCAR data using this procedure, important information concerning the transport spin-polarization may be obtained on candidate materials for applications of spintronics devices. We have proven that both ballistic and diffusive models yield essentially the same values of the spin polarization (with the accuracy of approximately 3%) practically within the full range of $P$. We have also shown that in some cases the observed correlation between $P_c$ and $Z$ can be due solely to systematic errors in the data analysis. At the same time we have confirmed a previously observed correlation for $P_c$ ($Z$) dependence in $CrO_2$, and some other material systems, in which case the interpolation to $Z = 0$ is legitimate. At the same time we conclude that, as of now, there is no extrapolation formula that is significantly better than the others. We have also noted that if all available PCAR data correspond to sizeable $Z$ in the ballistic model, the ballistic conditions should be independently verified before extrapolating to $Z = 0$. Much more work is needed to explain the mechanisms as to why the intrinsic value of the spin-polarization decreases when $Z$ increases when analyzing PCAR spectra using either limit. It is encouraging, however, that, the modified BTK formalism seems to be able to absorb a number of physical effects well beyond the scope of the underlying model into a single number, $Z$. Therefore, the values for the interfacial spin polarizations appear substantially more reliable than one could have anticipated from purely theoretical viewpoint.

The work at NRL was supported by the Office of Naval Research. B. N. acknowledges support by DARPA through ONR N00014-02-1-0886 and NSF Career grants.

# FIG CAPTIONS

FIG. 1 Comparison between the assigned values of the spin-polarization and Z parameters using the diffusive model ($P_d$, $Z_d$) and the fitted values using the ballistic model ($P_b$, $Z_b$). The two vertical axes show the shifts in (a) $P_b$ - $P_d$ and (b) $Z_b$ - $Z_d$. The highest percent shift for the polarizations is 5 %.

FIG. 2. Plot of $\Delta$ vs. $R_s$ from theoretically generated curves. Each curve was generated using fixed values of $T$, $R_s$, and $Z$ but with different $\Delta$'s and then fitted to obtain new values of $R_s$. The inset shows some of the curves generated and the new fitted values of $R_s$.

FIG. 3. $P_c$ vs. $\Delta$ for one point-contact spectrum of *Sn/LSMO*. The inset serves to illustrate that, in each extracted value of $P_c$ obtained from the models, a quality fit was achieved. All fits were done in the ballistic limit.

FIG. 4 Analyzed $G(V)$ curves of two point-contact spectra of *Pb/CrO$_2$* in the (a) ballistic limit and (b) the diffusive limit for positive bias voltage. The temperature used in the fits was T = 1.75 K, the value of $\Delta$ = 1.2 meV and $R_s$ ~ 0.5. The negative bias voltage spectra were symmetric to the positive bias spectra in all cases.

FIG 5 $P_c$ vs. $Z$ for *Pb/CrO$_2$* in the ballistic (filled squares) and the diffusive (empty circles) for several point-contacts. Extrapolations to Z = 0 linearly and quadratically give $P_c$ (Z=0) = 1.13 ± 0.06 and $P_c$ (Z = 0) = 1.05 ± 0.29 respectively. The value for the diffusive model with no extrapolation yields $P_c$ (Z = 0) ~ 0.85 ± 0.03. Dashed lines connect the two results from the same point contacts.

FIG. 6 Plots of $\chi^2$ for a linear extrapolation ($\chi_L$) and a quadratic one ($\chi_Q$). The results show that statistically these extrapolations are equivalent.

Table 1. Components of the modified BTK formalism. The following notations are used: $\beta = eV/\sqrt{|e^2V^2 - \Delta^2|}$, $F(x) = \cosh^{-1}(2Z^2 + x)/\sqrt{(2Z^2 + x)^2 - 1}$.

|  | $eV < \Delta$ | $eV < \Delta$ |
|---|---|---|
| Ballistic non-magnetic | $\dfrac{2(1+\beta^2)}{\beta^2 + (1+2Z^2)^2}$ | $\dfrac{2\beta}{1+\beta+2Z^2}$ |
| Ballistic half-metallic | 0 | $\dfrac{4\beta}{(1+\beta)^2 + 4Z^2}$ |
| Diffusive non-magnetic | $\dfrac{(1+\beta^2)}{2\beta}\text{Im}[F(-i\beta) - F(i\beta)]$ | $\beta F(\beta)$ |
| Diffusive half-metallic | 0 | $\beta F[(1+\beta)^2/2 - 1]$ |

Table II. Fitted values from Eq. 1 for several ferromagnetic materials including **CrO₂** from this work. Also included are statistical comparisons of the linear ($\chi_L^2$), quadratic, ($\chi_Q^2$) and exponential ($\chi_E^2$) extrapolations.

| Material | $P_0$ | $\alpha\psi$ | $\chi_L^2$ | $\chi_Q^2$ | $\chi_E^2$ |
|---|---|---|---|---|---|
| *CrO₂* (this work) | 0.93 ± 0.03 | 0.245 ± 0.05 | 35.6 | 45.1 | 36.5 |
| *CrO₂* (Ref. 16) | 0.96 ± 0.02 | 1.5 ± 0.23 | 19.8 | 9.6 | 9.8 |
| *SRO* (Ref. 22) | 0.58 ± 0.01 | 0.59 ± 0.2 | 1.8 | 5.2 | 1.9 |
| *SRO* (Ref. 17) | 0.53 ± 0.01 | 1.12 ± 0.12 | 2.6 | 1.5 | 1.3 |
| *LSMO* (x = 0.4) (Ref. 10) | 0.82 ± 0.02 | 0.31 ± 0.03 | 10.8 | 6.8 | 7.7 |
| *LSMO* (x = 0.3) (Ref. 10) | 0.78 ± 0.01 | 0.243 ± 0.03 | 10.0 | 6.3 | 3.9 |
| *Ni* (Ref. 16) | 0.38 ± 0.01 | 1.94 ± 0.18 | 1.1 | 2.0 | 0.9 |

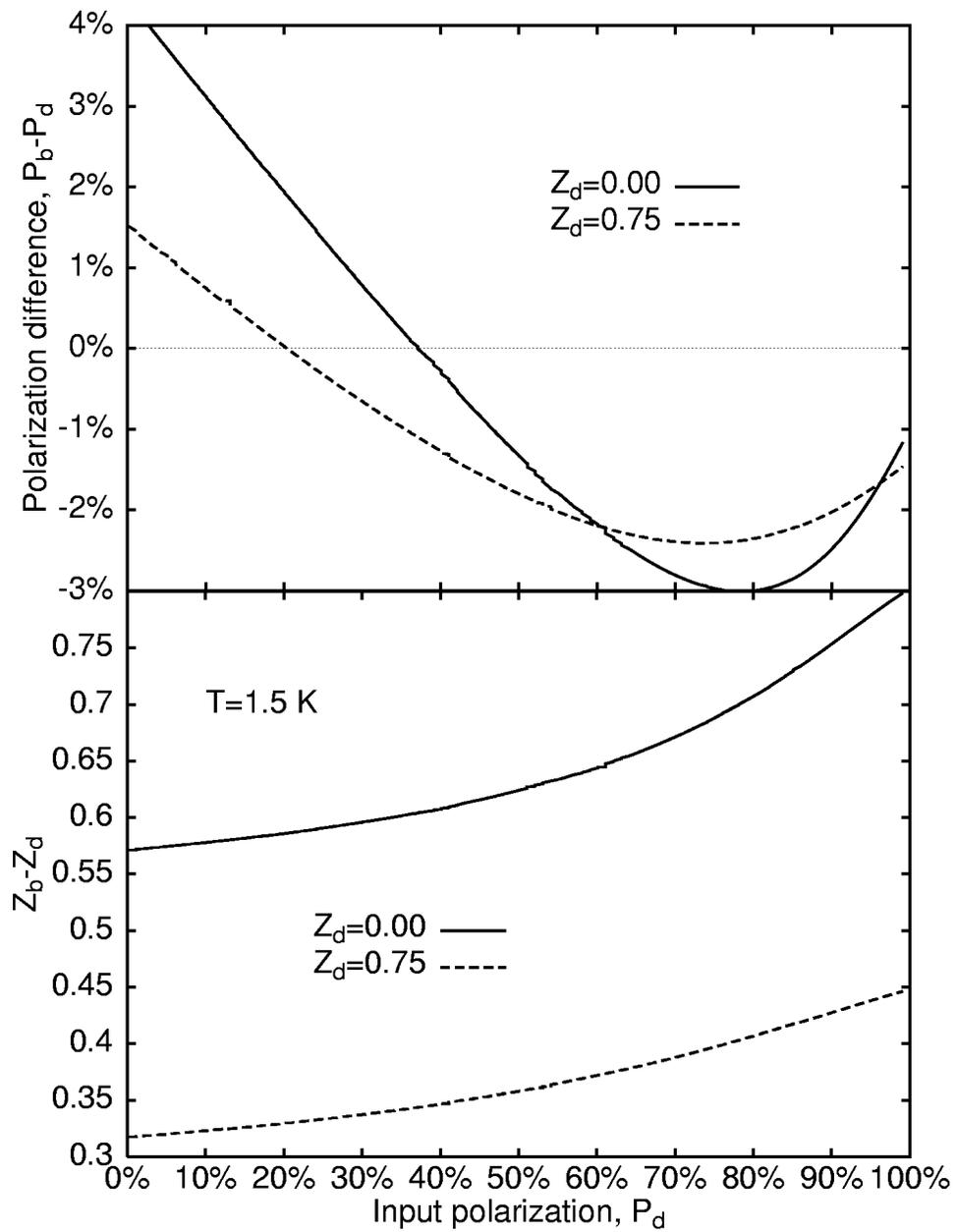

# FIG.1

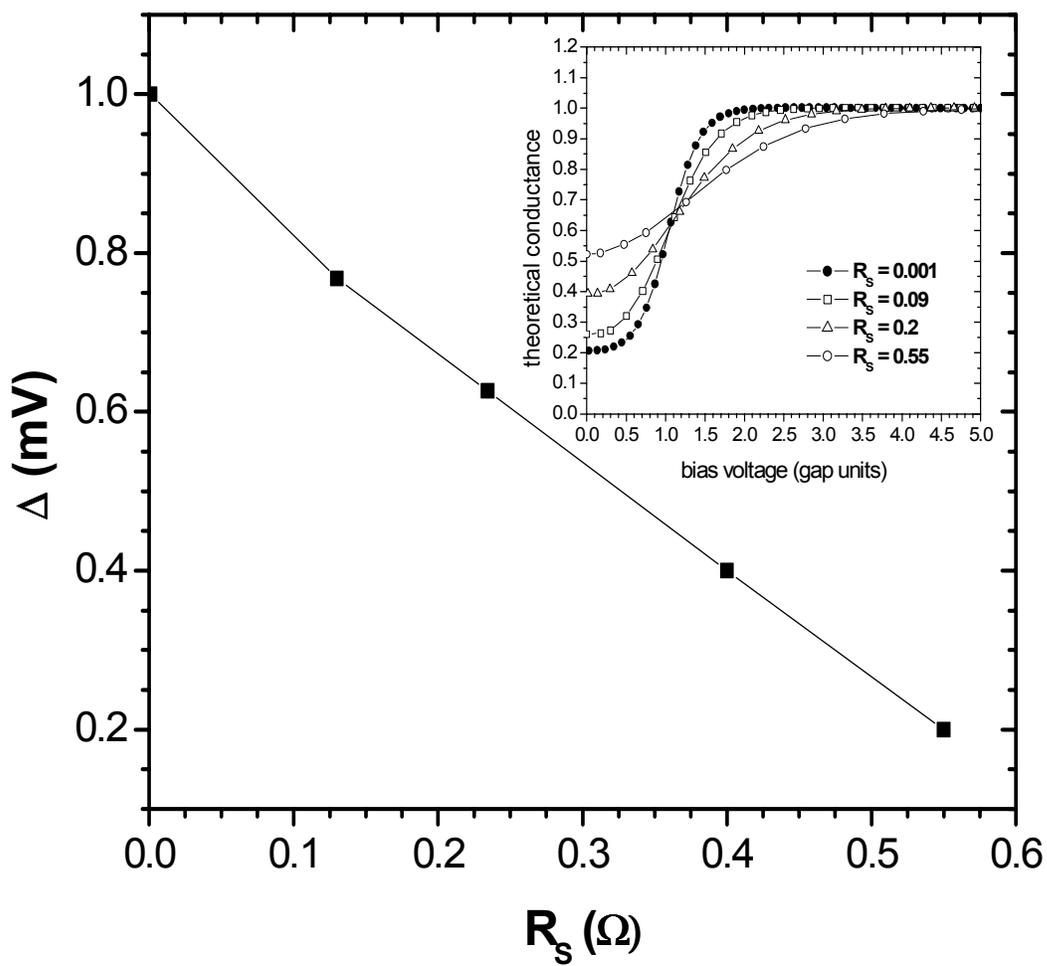

# FIG. 2

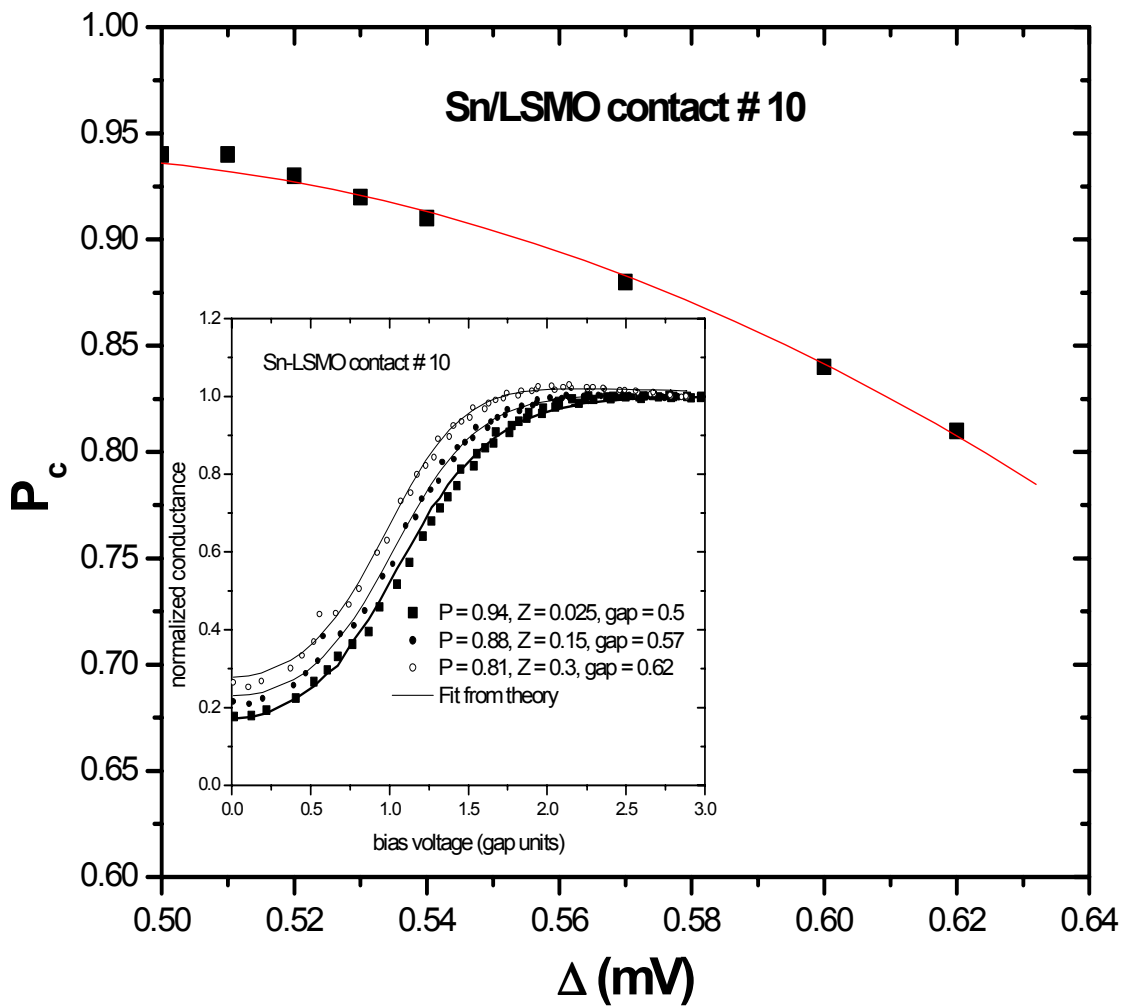

**FIG. 3**

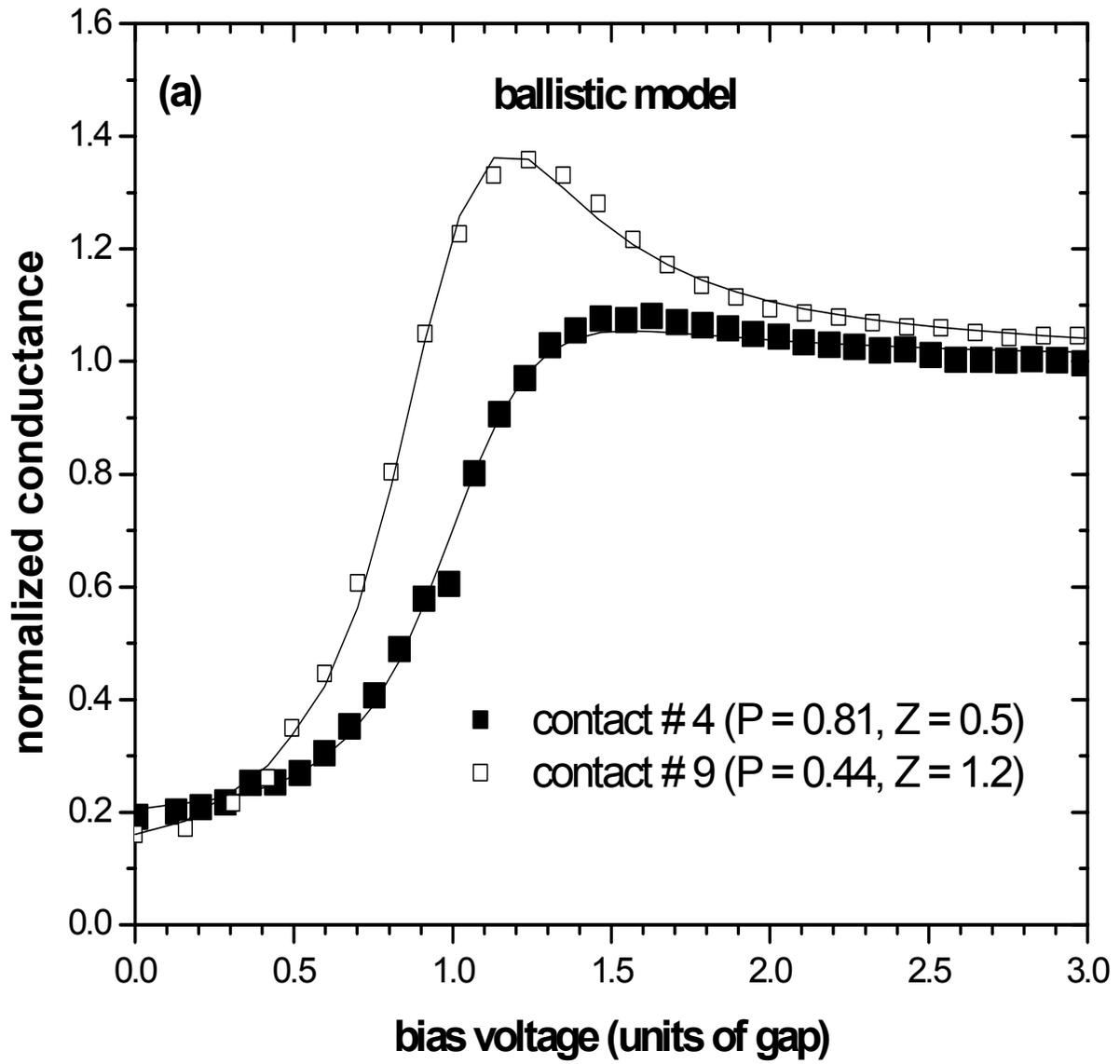

FIG. 4a

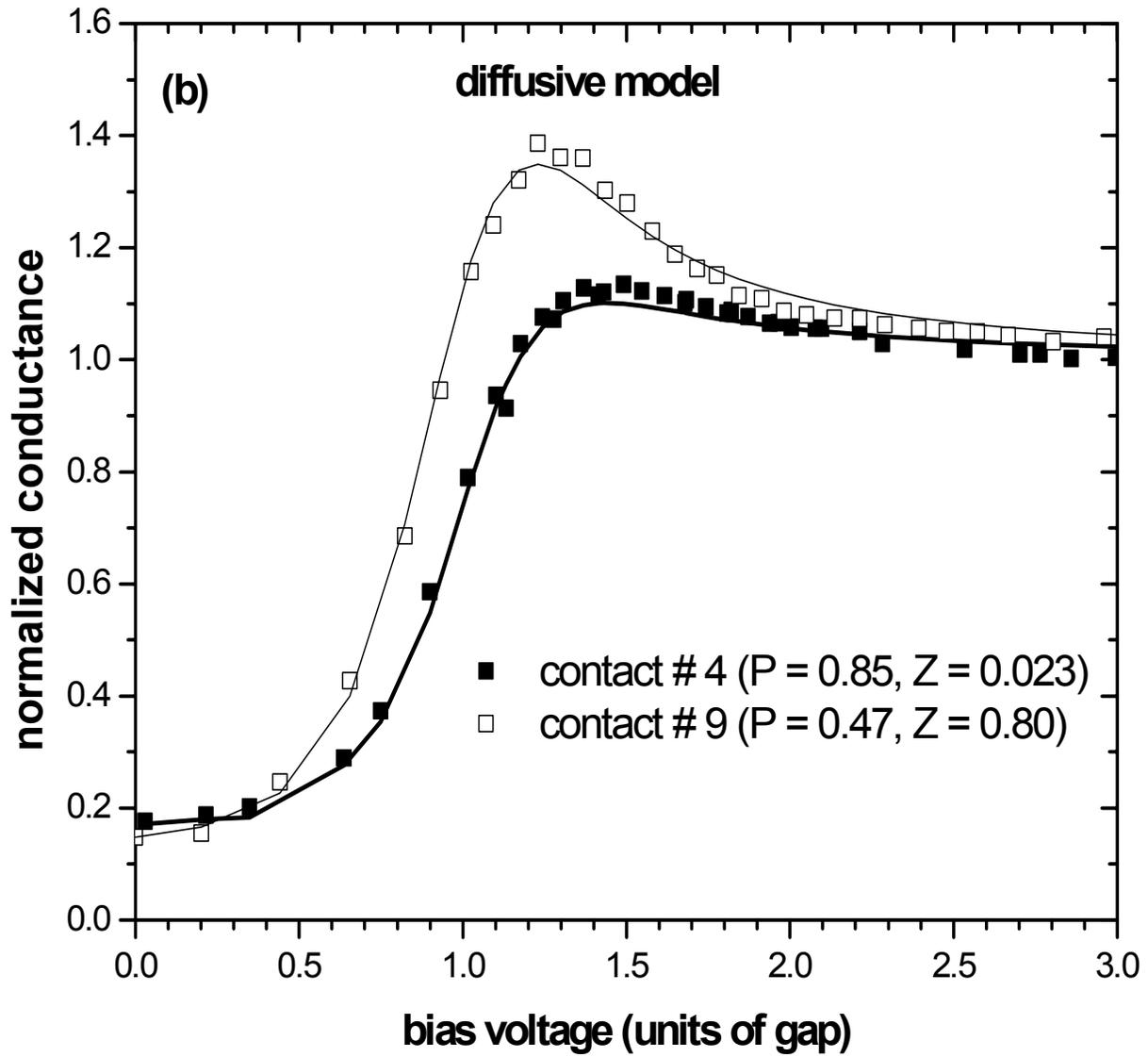

**FIG. 4b**

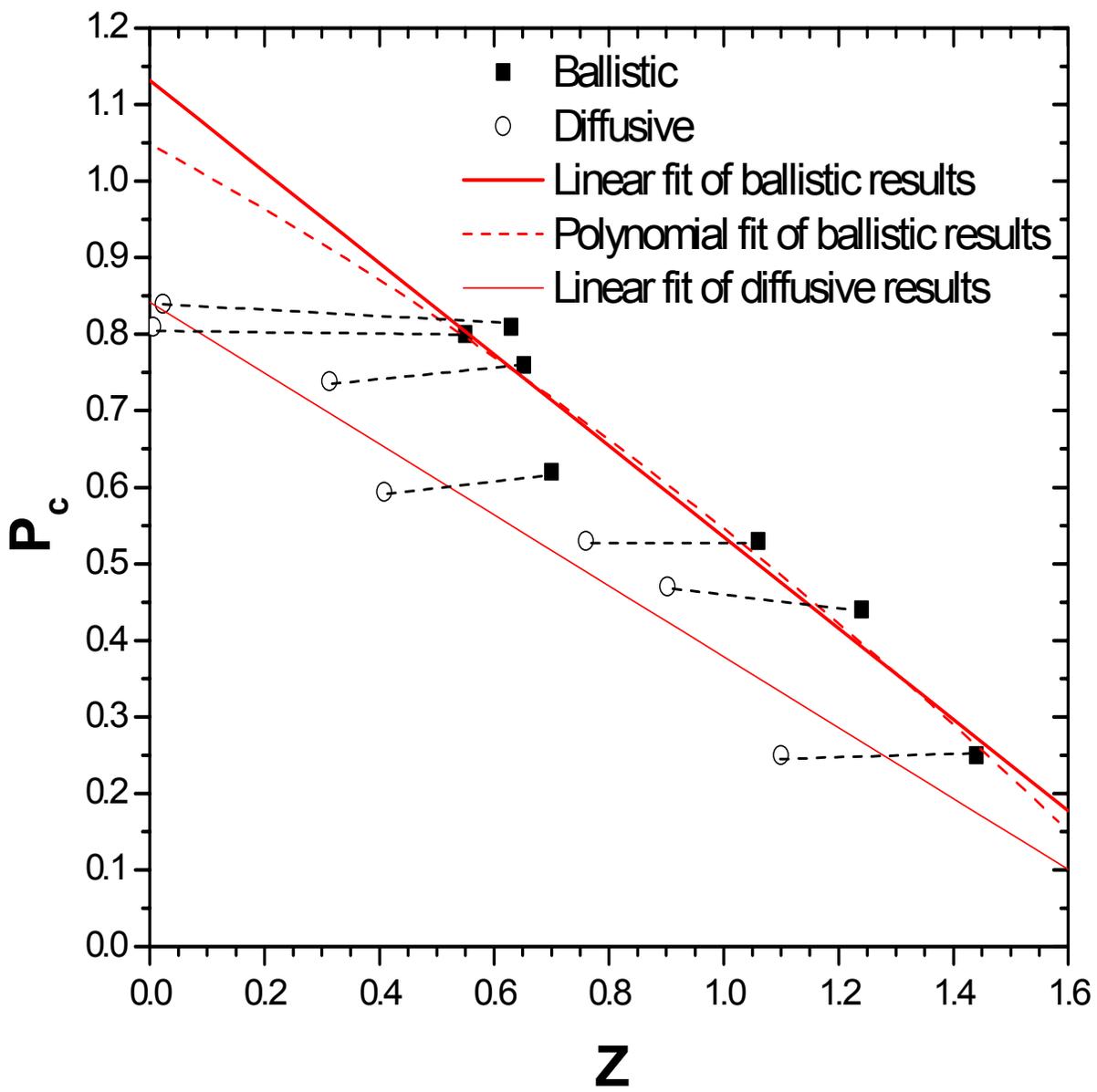

**FIG. 5**

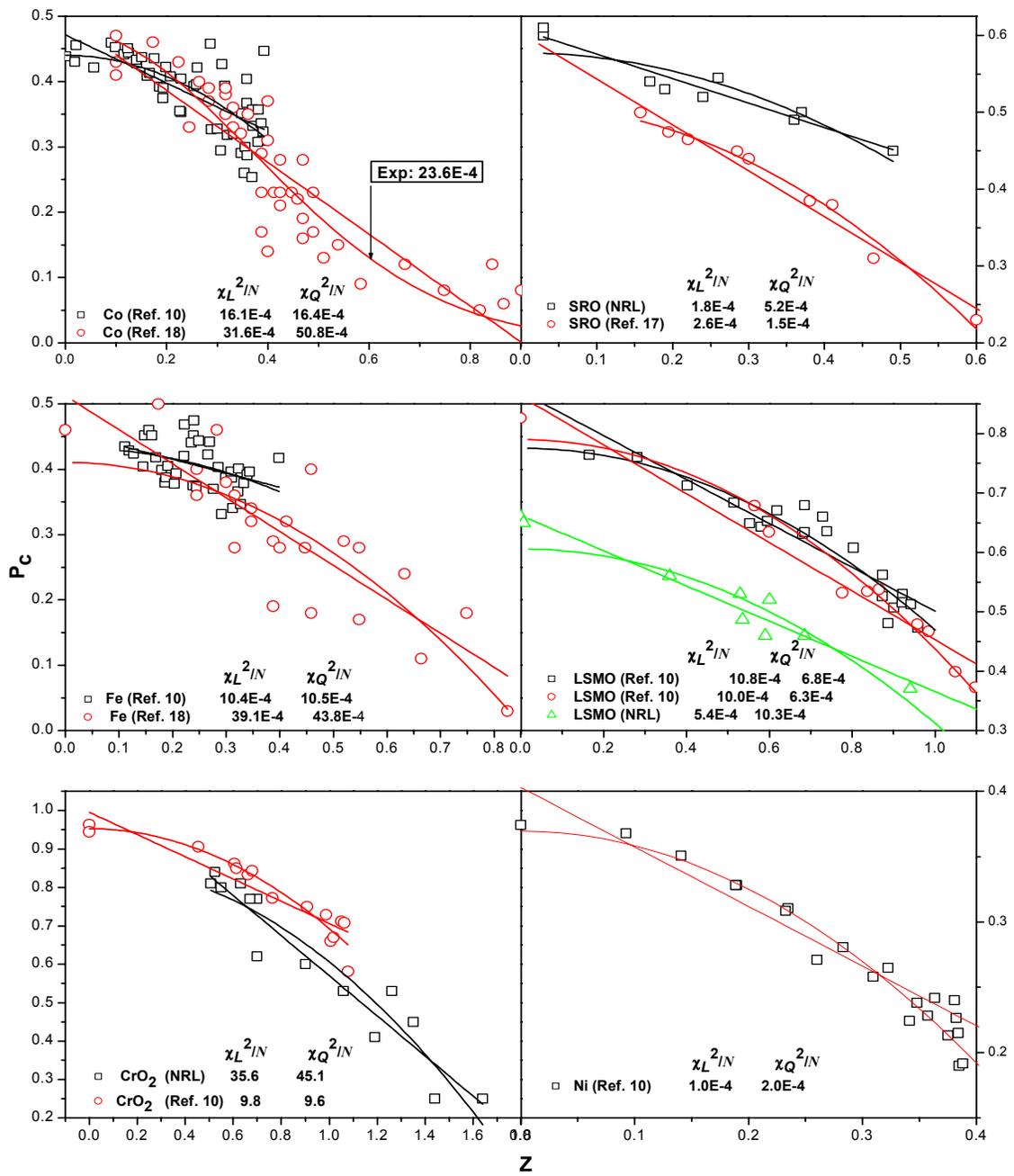

FIG. 6